\documentclass{sig-alternate}

\usepackage{fontenc}
\usepackage{scrextend}
\usepackage{amsmath}
\usepackage{amssymb}
\usepackage{fancybox}
\usepackage[font={footnotesize},skip=2pt]{caption}
\usepackage[font=footnotesize,captionskip=0pt,farskip=0pt,position=below]{subfig}
\usepackage{color}
\usepackage{colortbl}
\usepackage{array}
\usepackage{cite}
\usepackage{multirow}
\usepackage{multicol}
\usepackage{listings}
\usepackage{graphicx}
\usepackage{subfloat}
\usepackage{cite}
\usepackage{boxedminipage}
\usepackage[draft]{hyperref}
\usepackage[ruled,vlined,lined,commentsnumbered]{algorithm2e}
\usepackage{balance}
\usepackage{csvsimple}
\usepackage{longtable}
\usepackage[leftno,noindent]{lgrind}
\usepackage{booktabs}
\usepackage[font=small,labelfont=bf]{caption}

\usepackage{url}

\usepackage{enumitem}
\setlist{noitemsep} 

\makeatletter
\def\@copyrightspace{\relax}
\makeatother

\definecolor{yellow}{RGB}{255,255,153}
\definecolor{grey}{RGB}{224,224,224}
\newcommand{\highlight}{\cellcolor{grey}}

\newboolean{showcomments}
\setboolean{showcomments}{true}
\ifthenelse{\boolean{showcomments}}
 { \newcommand{\mynote}[2]{
      \fbox{\bfseries\sffamily\scriptsize#1}
        {\small$\blacktriangleright$\textsf{\emph{#2}}$\blacktriangleleft$}}}
        { \newcommand{\mynote}[2]{}}

{\medskip\vspace{0.5cm}
\noindent
\let\emph=\textbf
\begin{boxedminipage}{\columnwidth}\begin{center}\em}%
{\end{center}\end{boxedminipage}\vspace{0.5cm}\medskip}

\definecolor{DarkOrange}{rgb}{0.8,0.3,0.0}

\newcommand{\numSubjects}{10\xspace}
\newcommand{\numChanges}{48,272\xspace}

\newcommand{\precisionwild}{86.8\%\xspace}
\newcommand{\recallwild}{74\%\xspace}
\newcommand{\fmeasurewild}{80.4\%\xspace}

\begin{document}

\title{Watch out for This Commit!\\ A Study of Influential Software Changes}

\numberofauthors{6} 
%
\author{
\alignauthor
	Daoyuan Li\\
	\affaddr{University of Luxembourg}\\
	\email{daoyuan.li@uni.lu}
\alignauthor
	Li Li\\
	\affaddr{University of Luxembourg}\\
	\email{li.li@uni.lu}
\alignauthor
	Dongsun Kim\\
	\affaddr{University of Luxembourg}\\
	\email{dongsun.kim@uni.lu}
\and
\alignauthor
	Tegawend\'e F. Bissyand\'e\\
	\affaddr{University of Luxembourg}\\
	\email{tegawende.bissyande@uni.lu}
\alignauthor
	David Lo\\
	\affaddr{Singapore Management Univ}\\
	\email{davidlo@smu.edu.sg}
\alignauthor
Yves Le Traon\\
       \affaddr{University of Luxembourg}\\
       \email{yves.letraon@uni.lu}
}%
\date{30 Jan 2015}%

\maketitle%

\begin{abstract}
One single code change can significantly influence a wide range of software systems and their users.
For example, 1) adding a new feature can spread defects in several modules,
while 2) changing an API method can improve the performance of all client programs.
Developers often may not clearly know whether their or others'
changes are influential at commit time. Rather, it turns out to be
influential after affecting many aspects of a system later.

This paper investigates influential software changes and proposes an approach
to identify them early, i.e., immediately when they are applied.
We first conduct a post-mortem analysis to discover existing influential
changes by using intuitions such as \emph{isolated changes} and \emph{changes
referred by other changes} in \numSubjects open source projects.
Then we re-categorize all identified changes through an open-card sorting process. Subsequently, we conduct a survey
with 89 developers to confirm our influential change categories.
Finally, from our ground truth we extract features, including metrics such as
the complexity of changes, terms in commit logs and file centrality in
co-change graphs, to build machine learning classifiers.
The experiment results show that our prediction model achieves overall with random samples
\precisionwild precision, \recallwild recall and \fmeasurewild F-measure respectively.

\end{abstract}




\section{Introduction}
\label{sec:intro}

Current development practices heavily rely on version control systems to
record and keep track of changes committed in project repositories. While many
of the changes may be simply cosmetic or provide minor improvements, others
have a wide and long-term influence to the entire system and related systems.
Brudaru and Zeller~\cite{brudaru_what_2008} first illustrated examples of 
changes with long term-influence: 1) changing access privilege
(i.e., \texttt{private} $\rightarrow$ \texttt{public}), 2) changing kernel
lock mechanism, and 3) forgetting to check a null return. If we can predict
whether an incoming software change is influential or not, either positively
or negatively, just after it is committed, it could significantly improve
maintenance tasks (e.g., easing debugging if a new test harness is added) and provide insights for recommendation systems (e.g., code reviewers can focus on fewer changes).

The influence of a software change can however be hard to detect immediately since it often does
not involve immediate effects to other software elements. Instead, it can
constantly affect a large number of aspects in the software over time. Indeed,
a software change can be influential not only inside and/or beyond the project
repository (e.g., new defects in code base and new API calls from other
programs), but also immediately and/or  long after the changes have been
applied. The
following are examples of such influential changes:

{\bf Adding a new lock mechanism}: {\tt mutex-lock} features were introduced in 
Linux 2.6 to improve the safe execution of kernel critical code sections. However,
after their introduction, the defect density of Linux suddenly increased for 
several years, largely contributed by erroneous usage of these features.
Thus, the influence of the change was not limited to a specific set of modules.
Rather, it was a system-wide problem. 

{\bf Changing build configurations}: A small change in configuration files may
influence the entire program. In \texttt{Spring-framework}, a developer missed
file inclusion options when migrating to a new build system
(\texttt{$\ast$.aj} files were missing in \texttt{build.gradle}). 
This makes an impact since programs depending on the framework failed occasionally
to work. The reason of this failure (missed file) was hard to pinpoint.

{\bf Improving performance for a specific environment}:
\texttt{FastMath.floor()} method in \emph{Apache Commons Math} had a problem
with Android applications since it has a static code block that makes an
application hang about five seconds at the first call. Fixing this issue
improves the performance of all applications using the library. 


Unfortunately, existing techniques are limited to revealing the \emph{short-
term impact} of a certain software change. The short-term impact indicates an
immediate effect such as test case failure or coverage deviation. For example,
dynamic change analysis
techniques~\cite{ren_chianti:_2004,zhang_faulttracer:_2012} leverage coverage
metrics after running test cases. Differentiating coverage information
before/after making a change shows how the change influences other program
elements. Other approaches are based on similarity distances~\cite{robillard_retrieving_2008,sherriff_empirical_2008}.
 These
firstly identify clusters of program elements frequently changed
together or tightly coupled by analyzing revision histories. Then, they
attempt to figure out the best-matching clusters for a given change.
Developers can assume that program elements (e.g., files or methods) in the
cluster may be affected by the given change. Finally, change genealogy~\cite{herzig_capturing_2010,herzig_mining_2011,herzig_predicting_2013} 
approaches keep track of dependencies between subsequent changes, and can capture some long-term impact of changes.
However, it is limited to identifying source code entities and defect density. Overall, all  the above
techniques may not be successful in predicting a wide and long-term influence
of software changes. This was unfortunately inevitable since those existing
techniques focus only on explicit dependencies such as method calls.

\paragraph*{\bf Study Research Questions} In this study we are interested in investigating the following research questions:
\begin{itemize}
\itemsep0em
	\item[RQ1:] What constitutes an influential software change? Are there developer-approved definitions/descriptions of influential software changes?
	\item[RQ2:] What metrics can be used to collect examples of influential software changes?
	\item[RQ3:] Can we build a prediction model to identify influential software changes immediately after they are applied?
\end{itemize}


To automatically figure out whether an incoming software change is influential, we
designed a prediction technique based on machine learning classification. Since
the technique requires labeled training instances, we first discovered existing
influential changes in several open source projects in order to obtain baseline
data. Specifically, we collected \numChanges code commits from \numSubjects open source projects and did
post-mortem analysis to identify influential changes. This analysis examined
several aspects of influential changes such as controversial changes and breaking
behaviors. In addition, we manually analyzed whether those changes actually have
long-term influence to revision histories.
As a result, we could discover several influential changes from each subject.
We further label these changes to build category definition for influential software changes
through an open-card sorting process. These categories are then validated by
developers with experience in code review.   

Based on the influential changes we discovered in the above study, we extracted
feature vectors for machine-learning classification. These features include program structural
metrics~\cite{kim_classifying_2008}, terms in change
logs~\cite{kim_classifying_2008}, and co-change metrics~\cite{Beyer05}.
Then, we built a prediction model by leveraging machine learning algorithms such as
Na\"{i}ve Bayes~\cite{lewis:ecml:1998,mlbook} and Random
Forest~\cite{breiman_random_2001}.
To evaluate the effectiveness of this technique, we conducted experiments that
applied the technique to \numSubjects projects.
Experimental assessment results with a representative, randomly sampled, subset of our data show that 
our prediction model achieves overall
\precisionwild precision, \recallwild recall, and \fmeasurewild F-measure performance.

This paper makes the following contributions:

\begin{itemize}
\itemsep0em 
	\item Collection of influential software changes in popular open source projects.
	\item Definition of influential software change categories approved by the software development community.
	\item Correlation analysis of several program metrics and influential software
changes.
	\item Accurate machine-learning prediction model for influential software changes.
\end{itemize}

The remainder of this paper is organized as follows. After describing motivating
examples in Section~\ref{sec:motivation}, we present our study results of
post-mortem analysis for discovering influential changes in
Section~\ref{sec:preliminary}.
Section~\ref{sec:model} provides our design of a prediction model for
influential changes together with a list of features extracted from software changes.
In addition, the section reports the evaluation result of experiments in which
we applied the prediction model to open source projects.
Section~\ref{sec:discussion} discusses the limitations of our work.
After surveying the related work in Section~\ref{sec:related}, 
we conclude with directions for future research in
Section~\ref{sec:conclusion}.

\section{Motivating Examples}
\label{sec:motivation}


In the development course of software project, developers regularly commit
changes to project repositories. While some of those changes may simply be
cosmetic, a few others may be somehow influential not only inside and/or beyond
the repositories but also immediately and/or long after they are applied.
An influential software change can be recognized as such for various reasons, not
all of which are known while the change is being performed.

To motivate our study, we consider influential change 
examples identified from the Linux kernel project.
Linux is an appropriate subject as several changes in the kernel have
been influential. These changes are already highlighted in the literature~\cite{Palix10Faults,padioleau08} as
their long-term impact started to be noticed.
In this section, we present four different examples of influential changes in
Linux kernel and their impact.

\subsection{Collateral Evolution}
In the Linux kernel, since driver code, which makes up over 70\% of the source
code, is heavily dependent on the rest of the OS, any change in the interfaces
exported by the kernel and driver support libraries can trigger a large number
of adjustments in the dependent drivers~\cite{padioleau06}. 

Such adjustments,
known as collateral evolution, can unfortunately be challenging to implement
correctly. Starting with Linux 2.5.4, the USB library function {\em
usb\_submit\_urb} (which implements message passing) takes a second argument for
explicitly specifying the context (which was previously inferred in the
function definition). The argument can take one of three values: GFP\_KERNEL (no
constraints), GFP\_ATOMIC (blocking is not allowed), or GFP\_NOIO (blocking is
allowed but not I/O)). Developers using this USB library must then parse their
own code to understand which context it should be as in the example of Figure~\ref{listing:usb}.

This leads to bugs that keep occurring.
A study by Pallix {\em et al.}~\cite{Palix10Faults} has reported that, due to
the complexity of the conditions governing the choice of the new argument for
{\em usb\_submit\_urb}, 71 of the 158 calls to this function were initially
transformed incorrectly to use GFP\_KERNEL instead of GFP\_ATOMIC.

This change is interesting and constantly influential to a large portion of the
kernel, as its real impact could only be predicted if the analysis took into
account the semantics of the change. However, the extent of influences made by
the change is difficult to detect immediately after
the commit time since existing
techniques~\cite{ren_chianti:_2004,zhang_faulttracer:_2012,robillard_retrieving_2008,sherriff_empirical_2008}
focus only on the short-term impact.

\begin{figure}[!h]
\centering
{\parbox{\linewidth}{
\begin{lstlisting}[linewidth={\linewidth},frame=tb]
spin_lock_irqsave(&as->lock, flags);
if (!usbin_retire_desc(u, urb) &&
    u->flags & FLG_RUNNING &&
    !usbin_prepare_desc(u, urb) &&
-    (suret = usb_submit_urb(urb)) == 0) {
+    (suret = usb_submit_urb(urb, GFP_ATOMIC)) == 0) {
  u->flags |= mask;
} else {
  u->flags &= ~(mask | FLG_RUNNING);
  wake_up(&u->dma.wait);
  printk(KERN_DEBUG "...", suret);
}
spin_unlock_irqrestore(&as->lock, flags);
\end{lstlisting}
}}%
\caption{Code patch for adaption to the new definition of {\em usb\_submit\_urb}.
    In this case, when the API function is called, locks are held, so the programmer must use
    GFP\_ATOMIC to avoid blocking. Its influence was propagated to most
    drivers using this library and mostly resulted in defects.}
\label{listing:usb}
\vspace{-0.5cm}
\end{figure}

\subsection{Feature Replacement}
\label{sec:motivation2}

\begin{figure}[h!]
\includegraphics[width=\linewidth]{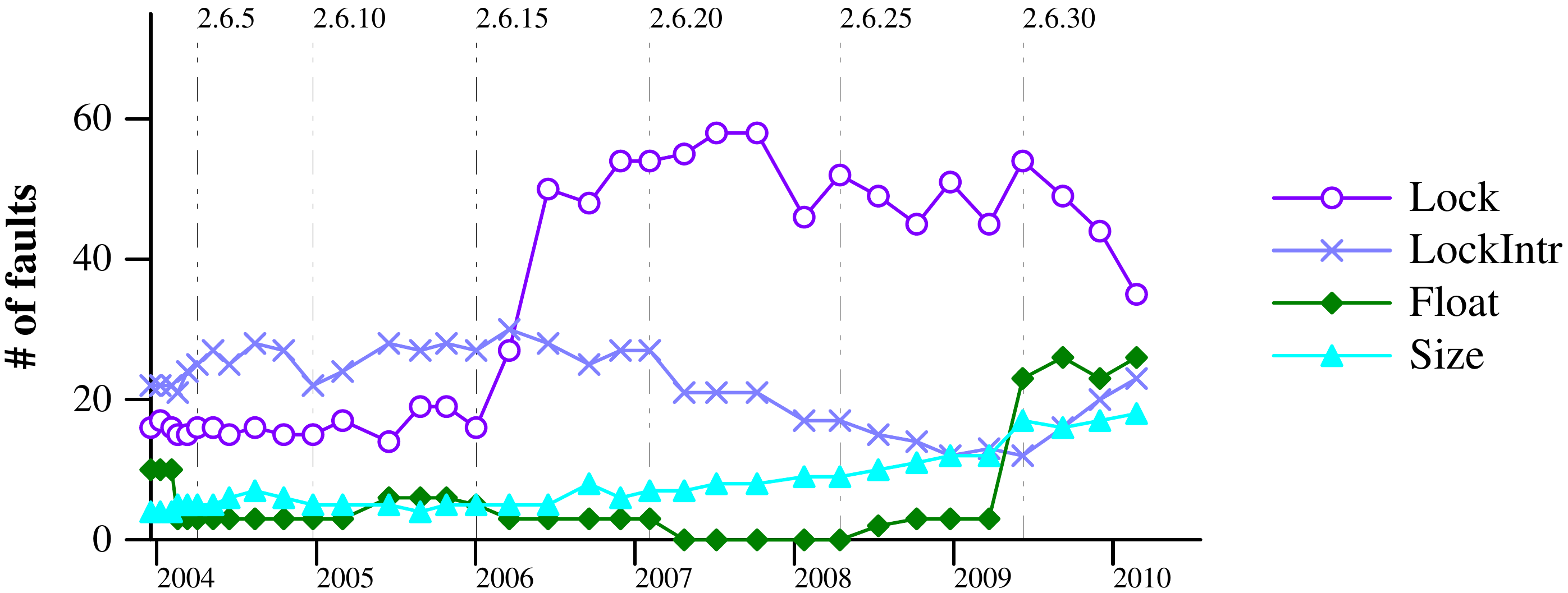} \caption{Evolution
of faults in Linux 2.6 kernel versions for {\em Lock}, {\em LockIntr}, {\em
Float} and {\em Size} fault categories~(see~\cite{Palix10Faults}). Faults
relevant to \emph{Lock} suddenly increased after Version 2.6.16 while other
types of faults gradually decreased. In the version, a feature for Lock was
replaced and it was influential to many of kernel functions.}
\label{fig:lock_rise}
\vspace{-0.5cm}
\end{figure}

In general, the number of entries in each fault category (e.g., {\em NULL} or {\em Lock})
decreases over
time in the Linux code base~\cite{Palix10Faults}.
In Linux 2.6, however, as illustrated in Figure~\ref{fig:lock_rise}, there are
some versions in which we can see a sudden rise in the number of faults.
This was the case of faults in the Lock\footnote{To avoid {\em Lock/LockIntr}
faults, release acquired locks, restore disable interrupts and do not double
acquire locks~\cite{Palix10Faults,Diagnosys}.
} 
category in Linux 2.6.16 due to a replacement of
functionality implementation.
In Linux 2.6.16, the functions {\em mutex\_lock} and {\em mutex\_unlock} were
introduced to replace mutex-like occurrences of the semaphore functions {\em
down} and {\em up}. The study of Palix {\em et al.} again revealed that 9 of the
11 Lock faults introduced in Linux  2.6.16 and 23 of the 25 Lock faults
introduced in Linux 2.6.17 were in the use of {\em mutex\_lock}.

If the replacement is identified earlier as an influential change to most of
kernel components (and other applications), it may prevent the defects from
recurring everywhere since the change is likely to be an API
change~\cite{linares-vasquez_api_2013,dig_how_2006}.
The developer who committed the new feature did not realize the influence and thus,
there was no early heads-up for other developers.


\subsection{Revolutionary Feature}

\begin{figure}[h!]%
    \centering
    \includegraphics[width=\linewidth]{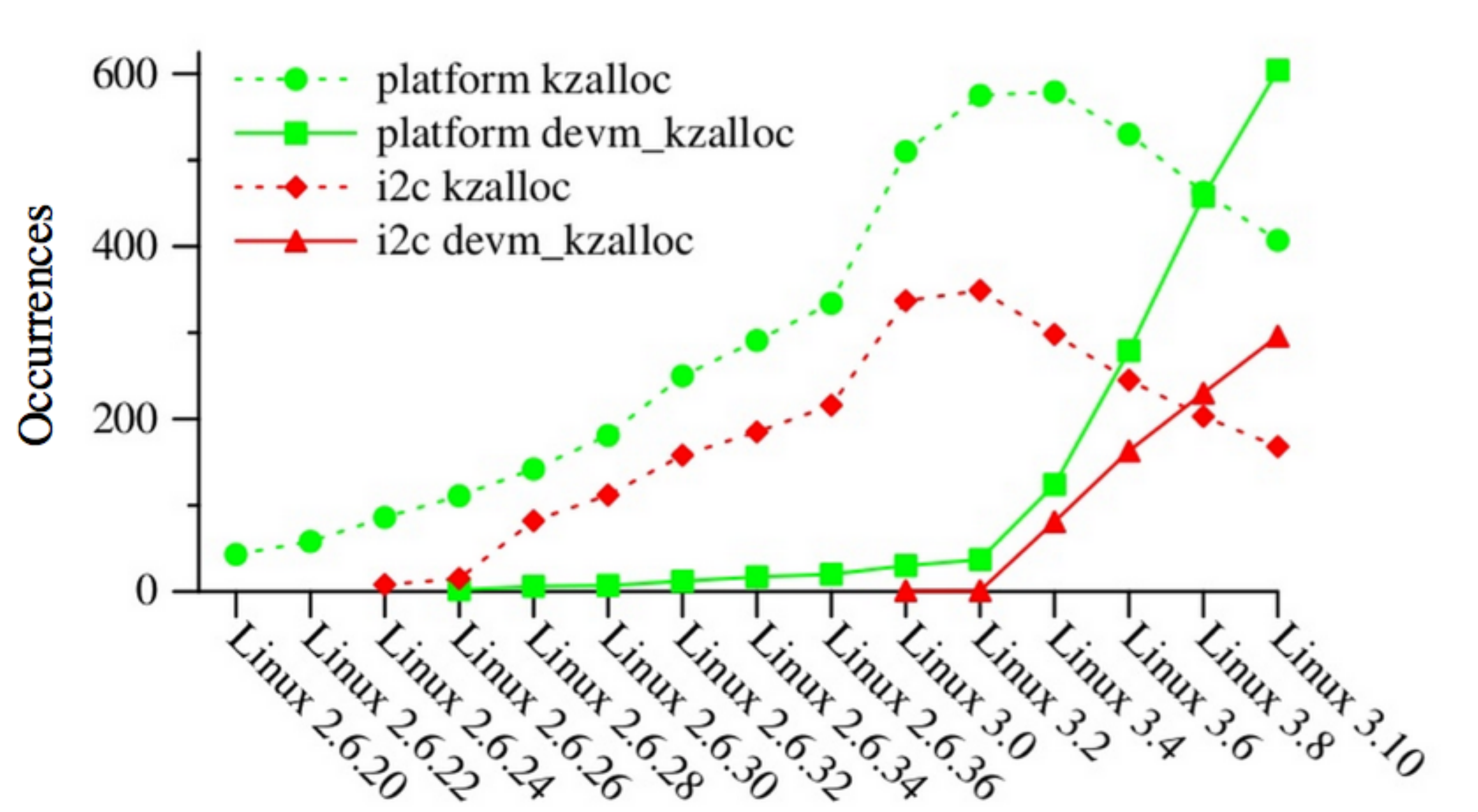}
    \caption{Usage of memory allocation primitives in Linux kernel (See~\cite{recipes_devm}). {\em kzalloc} is the traditional API for memory allocation,
before managed memory ({\em devm}) was introduced in Linux.}
    \label{fig:devm}
    \vspace{-0.5cm}
\end{figure}

An obvious influential change may consist in providing an implementation of
a completely new feature, e.g., in the form of an API function.
In the Linux kernel repository, Git commit {\tt 9ac7849e} introduced device resource management
API for device drivers. Known as the {\em devm} functions, the API
provides memory management primitives for replacing {\em kzalloc} functions.
This code change is a typical example of influential change with a long-term impact.
As depicted in Figure~\ref{fig:devm}, this change has first gone unnoticed
before more and more people started using {\em devm} instead of {\em kzalloc}.
Had the developers recognized this change as highly influential, {\em devm} could have been adopted earlier and result in less bugs and better performance in driver code.

\subsection{Fixes of Controversial/Popular Issues}
Some issues in software projects can be abnormally discussed or commented longer than others.
Code changes that fix them will be influential for the project. The characteristics of a controversial/popular issue
 is that its resolution is of interest for a large number of developers, and it takes more time
to resolve them than the average time-to-fix delay. Thus, we consider that an issue report
which is commented on average more than other issues and is fixed very long after it
is opened, is about a controversial/popular issue. In Linux, Git commit {\tt bfd36103}
resolved Bug \#16691 which remained unresolved in the bug tracking system for 9 months
and was commented about 150 times.

\begin{table*}[t!]
    \caption{Observational study subjects - Data reflect the state of repositories as of 26 January 2015}
    \label{tab:subjects}
\centering
\resizebox{.9\linewidth}{!}{
\begin{tabular}{rlccccc}
    Project Name & Description & \# Files & \# Commits & \# Developers & \# Issues & \# Resolved Issues \\
\toprule
{\tt Commons-codec} & General encoding/decoding algorithms & 635 & 1,424 & 24 & 195 & 177 \\
{\tt Commons-collections} & Extension of the Java Collections Framework& 2,983 & 2,722& 47 & 542 & 512\\
{\tt Commons-compress} & Library for working with file compression& 619 & 1,716& 24 & 308 & 272\\
{\tt Commons-csv} & Extension of the Java Collections Framework& 141 & 956& 18 & 147 & 119\\
{\tt Commons-io} & Collection of utilities for CSV file reading/writing & 631 & 1,718& 33 & 454 & 365\\
{\tt Commons-lang} & Extra-functionality for java.lang& 1,294 & 4,103& 46 & 1,073& 933\\
{\tt Commons-math} & Mathematics \& Statistics components & 4,582 & 5,496& 37 & 1,194& 1,085\\
{\tt Spring-framework} & Application framework for the Java platform& 19,721 & 9,748& 153& 3,500 & 2,632 \\
{\tt Storm} & Distributed real-time computation system& 2,038 & 3,534 & 189 & 637 & 321 \\
{\tt Wildfly} & aka JBoss Application Server& 31,699 &  16,855 & 307& 3,710 & 2,993\\
\midrule
{\tt Total} & & 64,388 & 48,272 & 878 & 11,760 & 9,409\\
\bottomrule
\end{tabular}
\vspace{-0.5cm}

}

    \vspace{-0.3cm}
\end{table*}

\section{Post-mortem Analysis for IC{\small \bf S}}
\label{sec:preliminary}
In this study, we focus on systematically discovering influential
changes.  Although the motivating examples described in
Section~\ref{sec:motivation} show some intuitions on influential
changes, it is necessary to reveal a larger view to figure out the
characteristics of these changes. Therefore, we collected
\numChanges changes from
\numSubjects popular open-source projects and conducted an observational study.


Since there are too many changes in software repositories and it is not
possible for us to inspect all, we get a set of changes that are likely to
have a higher density of influential changes. We are able to get this set by
leveraging several intuitions obtained from examples described in
Section~\ref{sec:motivation}.

The study design basically addressed three different criteria to
discover influential changes: 1) popular changes in the sense that they
have been somehow noticed by other developers and users,
2) anomalies in change behaviors, and 3) changes that are related to controversial/popular issues. 
These criteria are designed to conduct post-mortem analysis and represent how people
can recognize influential changes in hindsight.

For changes in these categories, we manually examine them using the following
procedure:
\begin{itemize}
	\item First of all, authors of this article ask themselves individually whether a change is really influential.
	They manually verify that the assumptions behind the specific criteria used to identify a change are supported.
	\item Then we cross-check the answers to reach an consensus among the authors.
\item Afterwards, we double check that these changes are really influential in
the eyes of developers by doing card sorting and surveying professional
developers.
\end{itemize}


\subsection{Data Collection}
The experiment subjects in this study are shown in 
Table~\ref{tab:subjects}. The \numSubjects popular
projects were considered since they have sufficient number of 
changes in their revision histories. In addition,
these projects stably maintained their issue tracking systems so that we
could keep track of how developers discussed to make software changes.

For each subject, we collected all available change data (patches and relevant
files information) as well as commit metadata (change date and author details) from
the source code repository.
Additionally, issue reports from the corresponding issue tracking system were
collected together. We further mined issue linking information from commit
messages and issue reports wherever possible: e.g., many commit messages
explicitly refer to the unique ID of the issue they are addressing, whether a
bug or a feature request.

\subsection{Systematic Analysis}
To systematically discover potential influential changes among the changes collected from
the subject projects, we propose to build on common intuitions about how a single change can
be influential in the development of a software project.

\subsubsection{Changes that address controversial/popular issues}
\label{sec:blocking}
In software projects, developers use issue tracking systems
to track and fix bugs and for planning
future improvements. When an issue is reported, developers and/or
users may provide insights of how the issue can be investigated. 
Attempts to resolve the issue are also often recorded in the issue tracking
system.

Since an issue tracking system appears as an important place to discuss
about software quality, we believe it is natural to assume that heated
discussions about a certain issue may suggest the importance of this
specific issue. Furthermore, when an issue is finally resolved after an exceptionally
lengthy discussion, all early fix attempts and the final commit that resolves
the issue should be considered to be influential. Indeed all these
software changes have contributed to close the discussion, unlock whatever
has been blocking attention from other issues, and satisfy the majority of
stakeholders. 

To identify controversial/popular issues in projects, we first searched for
issues with an overwhelmingly larger number of comments than others within the
same project. In this study, we regarded an issue as a controversial/popular issue if the
number of its comments is larger than the 99th percentile of issue comment
numbers. Applying this simple criteria, we could identify a set of issues that
 are controversial/popular. Afterwards, we collected all commits that were
associated to each of the controversial/popular issues and tag them as potentially
influential.


An example was found in Apache Math. An
issue\footnote{\small https://issues.apache.org/jira/browse/MATH-650} with 62
comments was detected by this analysis. This issue is about a simple glitch of
an API method; the API hangs 4--5 seconds at the first call on a specific
Android device. The corresponding
changes\footnote{Commits {\tt\small 52649fda4c9643afcc4f8cbf9f8527893fd129ba} and
{\tt\small 0e9a5f40f4602946a2d5b0efdc75817854486cd7}} fixed the glitch and
closed the issue.

To confirm that a change related to an identified controversial/popular issue (based on the number of comments) is truly influential,
we verify that 1) the discussion indeed was about a controversy and 2) the change is a key turning point in the discussion.
Table~\ref{tab:blocking}
compiles the statistics of changes linked to inferred controversial/popular issues as well as the
the number of influential changes manually confirmed among those changes.

\begin{table}[h!]
\centering
\caption{Statistics of identified influential changes related to controversial/popular issues.}
\centering
\resizebox{1\linewidth}{!}{
\begin{tabular}{lcc}
    Project Name & \# changes linked to controversial/popular issues & \# influential changes\\
\toprule

{\tt Commons-codec} & 26 & 3 \\
{\tt Commons-collections} & 12 & 8 \\
{\tt Commons-compress} & 7 & 4 \\
{\tt Commons-csv} & 5 & 5 \\
{\tt Commons-io} & 10 & 0 \\
{\tt Commons-lang} & 29 & 15 \\
{\tt Commons-math} & 38 & 8 \\
{\tt Spring-framework} & 53 & 42 \\
{\tt Storm} & 40 & 3 \\
{\tt Wildfly} & 20 & 18 \\
\midrule
Total & 240 & 106 \\
\bottomrule

\end{tabular}
}

\label{tab:blocking}
\vspace{-0.3cm}
\end{table}

\subsubsection{Anomalies in Change Behaviors}
\label{sec:isolated}

During software development, source code modifications
are generally made in a consecutive way following a somehow regular rhythm.
Break in change behaviors may thus signify abnormality and suggest
that a specific commit is relatively more important than others. 
For instance consider the following scenario: a certain file within a repository
after a period of regular edits remains unchanged for a period
of time, then is suddenly updated by a single change commit, and afterwards
remains again unchanged for a long time. Such a sudden and abnormal change 
suggests an urgency to address an issue, e.g., a major bug fix.
In our observational study we consider both break in behaviors in the edit rhythm
of each files and the edit rhythm of developers.
An anomaly in change behavior may be an out-of-norm change that developers do not notice,
 or a change to stable behavior that many developer/parts of code rely on.

\begin{figure}[t!]
\centering
\includegraphics[width=0.9\linewidth]{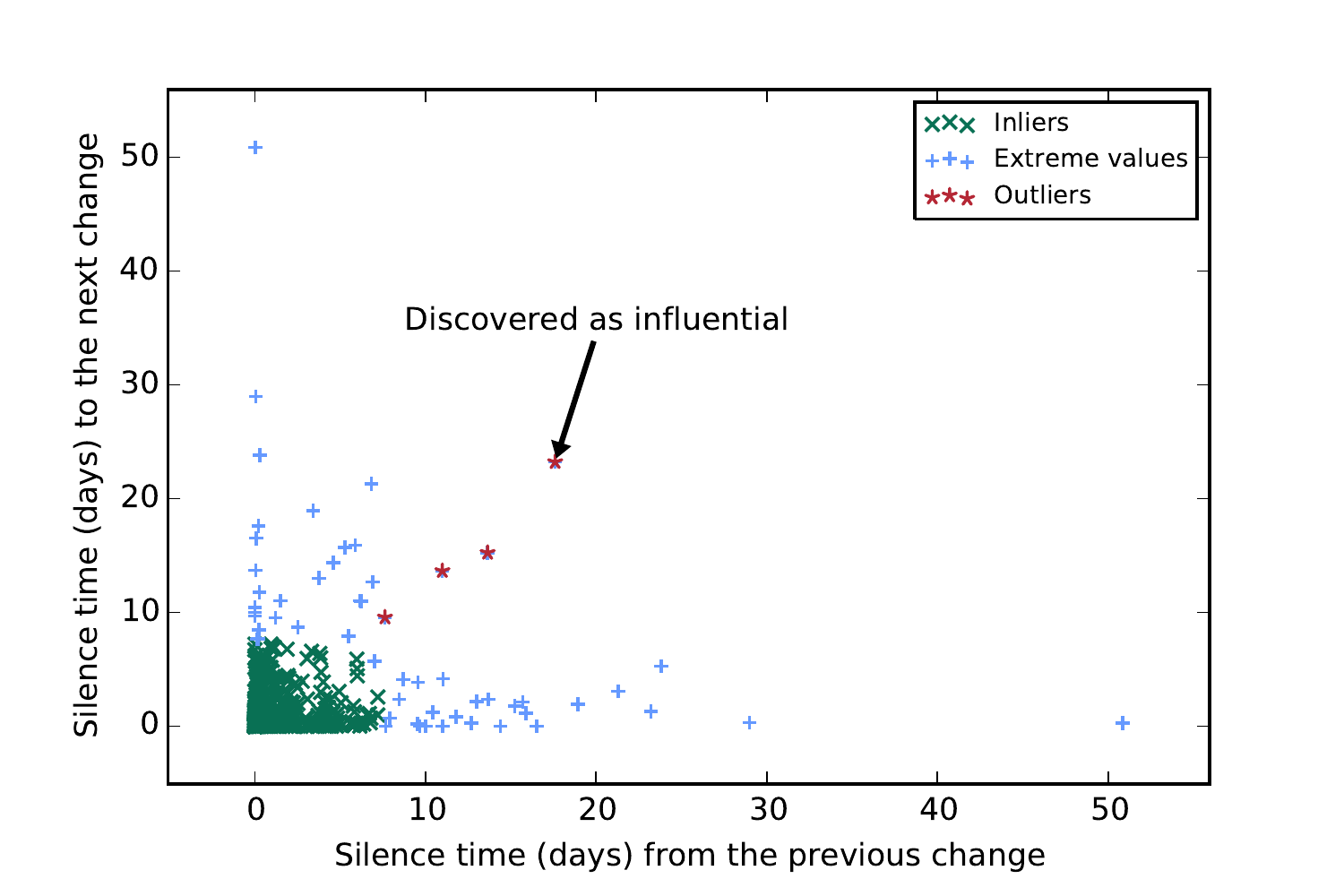}
\caption{Outlier detection to discover isolated commits for {\em build.gradle} file in {\tt Spring framework}.}
\label{fig:build-gradle}
\vspace{-0.5cm}
\end{figure}

In this study, for each file in the project we considered all commits that
modify the file. For each of those commits, we computed the time differences
from the previous commit and to the next commit. Then, we mapped these two time
lags to a two dimensional space and used Elliptic Envelope outlier detection~\cite{rousseeuw1999fast} to
identify ``isolated commits''.
In Figure~\ref{fig:build-gradle}, we can visualize the outliers discovered for
the changes on the {\em build.gradle} file from the {\tt Spring} project
subject.
The highlighted outlier represents a commit\footnote{Commit \tt\small
a681e574c3f732d3ac945a1dda4a640ce5514742} for including AspectJ files in the
Spring-sources jar file. This small commit is influential as it
fixes the build of Spring-framework.

Individual project contributors also often exhibit abnormal
behaviors which may suggest influential code changes.
For instance, one core developer constantly contributes
 to a specific project. If such a developer
submits isolated commits (i.e., commits that are a long time away from
the author's previous commit as well as his/her next commit), this
might be recognized as an emergency case where immediate attention is needed. 

In this study, we also systematically classified isolated commits based on
developer behaviors as potentially influential. For example, from commits by
developer {\em Stefan Bodewig} in {\tt Commons-COMPRESS}, we found an isolated
commit\footnote{Commit \tt\small fadbb4cc0e9ca11c371c87ce042fd596b13eb092}
where he proposed a major bug fix for the implementation of the {\em
ZipArchiveEntry} API. Before this influential software change, any attempt to create a
zip file with a large number of entries was producing a corrupted file.



To confirm that an isolated change is influential we verify that 1) its importance is clearly stated in the change log and
2) the implication of the change for dependent modules and client applications is apparent.
Table~\ref{tab:isolated} provides the statistics on detected isolated commits
and the results of our manual analysis on those  commits to confirm influential changes.

\begin{table}[h!]
\centering
\caption{Statistics of identified isolated commits and the associated manually confirmed influential changes.}

\centering
\resizebox{1\linewidth}{!}{
\begin{tabular}{lcc}
    Project Name & \# isolated commits & \# influential changes\\
\toprule

{\tt Commons-codec} & 7 & 3 \\
{\tt Commons-collections} & 28 & 9 \\
{\tt Commons-compress} & 17 & 5 \\
{\tt Commons-csv} & 13 & 4 \\
{\tt Commons-io} & 18 & 5 \\
{\tt Commons-lang} & 22 & 7 \\
{\tt Commons-math} & 29 & 5 \\
{\tt Spring-framework} & 56 & 8 \\
{\tt Storm} & 48 & 7 \\
{\tt Wildfly} & 213 & 1 \\
\midrule
Total & 451 & 54 \\
\bottomrule

\end{tabular}
}

\label{tab:isolated}
\vspace{-0.5cm}
\end{table}

\subsubsection{Changes referred to in other changes}
We considered that popular changes are potentially influential.
These are changes that other developers have somehow noticed (e.g., incomplete fix,
API change that causes collateral evolution). Indeed,
when developers submit software changes to a project, they usually submit
also a commit message introducing what their patch does. Occasionally,
developers refer to others' contributions in these messages. This kind of
behaviors suggests that the referred contribution is influential, at
least to a certain extent. For example, in the {\tt Commons-CSV} project, 
commit\footnote{Commit \tt\small 93089b260cd2030d69b3f7113ed643b9af1adcaa} 
{\tt 93089b26} is referred by another commit\footnote{Commit \tt\small
05b5c8ef488d5d230d665b9d488ca572bec5dc0c}.
This commit implemented the capability to detect start of line, 
which is surely an influential change for the implementation of CSV format reading.

Because some of the projects have switched from
using Subversion to using Git, we first managed to create a mapping between
the Subversion revision numbers (which remain as such in the commit messages) and
the newly attributed Git Hash code. 
To confirm that a change referenced by other changes is influential we verify that 1) it is indeed referenced by 
others because it was inducing their changes, and 2) the implication of the change for dependent modules and client applications
are apparent. Table~\ref{tab:referenced} provides the
statistics of influential changes derived with this metric.

\begin{table}[t!]
\centering
\caption{Statistics of identified referenced commits and influential commits.}
\centering
\resizebox{1\linewidth}{!}{
\begin{tabular}{lcc}
    Project Name & \# referenced commits & \# influential changes\\
\toprule

{\tt Commons-codec} & 8 & 3 \\
{\tt Commons-collections} & 3 & 1 \\
{\tt Commons-compress} & 3 & 0 \\
{\tt Commons-csv} & 3 & 2 \\
{\tt Commons-io} & 5 & 1 \\
{\tt Commons-lang} & 21 & 2 \\
{\tt Commons-math} & 43 & 3 \\
{\tt Spring-framework} & 1 & 1 \\
{\tt Storm} & 1 & 0 \\
{\tt Wildfly} & 11 & 9 \\
\midrule
Total & 99 & 22 \\
\bottomrule

\end{tabular}
}

\label{tab:referenced}
\vspace{-0.5cm}
\end{table}

\subsection{Qualitative Assessment Results}

We then set to assess the quality of the metrics used in our observational
study. We manually checked all potential influential changes yielded by the
systematic analysis. We further randomly pick change commits from each project
and manually check the percentage of changes that are influential. The
comparison between the two types of datasets aimed at validating our choices
of post-mortem metrics to easily collect influential changes.
Table~\ref{tab:qualitative} provides results of the qualitative assessment.
For each project, the random dataset size is fixed to 20 commits, leading to a
manual checking of 200 changes. Our systematic analysis findings produce
change datasets with highest rates of ``truly'' influential changes (an order
of magnitude more than what can be identified in random samples).

\begin{table}[h!] \centering \caption{Qualitative assessment results. We
compared the percentage of changes that were actually manually confirmed to be
influential from the datasets yielded by our systematic analysis and a random
selection in projects. Note that we count unique commits in this table, since 
some commits fall into more than one categories.}
\vspace{0.1cm}
\centering
\resizebox{1\linewidth}{!}{
\begin{tabular}{l c c c | c c}
  \multirow{2}{*}{\bf Project Name} & 
  \multicolumn{3}{c|}{\bf Systematic Analysis Findings} & 
  \multicolumn{2}{c}{\bf Random Selection} \\
  & Total & Influential & Rate & Influential & Rate \\

\toprule

{\tt Commons-codec} & 40 & 8 & 20.0\% & 0 & 0.0\% \\
{\tt Commons-collections} & 42 & 17 & 40.5\% & 0 & 0.0\%\\
{\tt Commons-compress} & 27 & 9 & 33.3\% & 1 & 5.0\% \\
{\tt Commons-csv} & 21 & 11 & 52.4\% & 1 & 5.0\% \\
{\tt Commons-io} & 33 & 6 & 18.2\% & 0 & 0.0\%\\
{\tt Commons-lang} & 72 & 24 & 33.3\% &0 & 0.0\%\\
{\tt Commons-math} & 108 & 14 & 13.0\% & 0 & 0.0\%\\
{\tt Spring-framework} & 110 & 51 & 46.4\% & 1 & 5.0\%\\
{\tt Storm} & 89 & 10 & 11.2\% & 1 & 5.0\% \\
{\tt Wildfly} & 243 & 27 & 11.1\% & 1 & 5.0\% \\
\midrule
Total & 785 & 177 & {\bf 22.5\%} & 5 & {\bf 2.5\%}\\
\bottomrule

\end{tabular}}

\label{tab:qualitative}
\vspace{-0.5cm}
\end{table}

{\bf Conclusion:}{\it 
The difference in influential rate values with random shows that our 
post-mortem metrics (isolated changes, popular commits, changes unlocking issues) are indeed 
good indicators for collecting \underline{some} influential software changes.
                }

\begin{figure*}[hbt!]
\centering
\includegraphics[width=\textwidth]{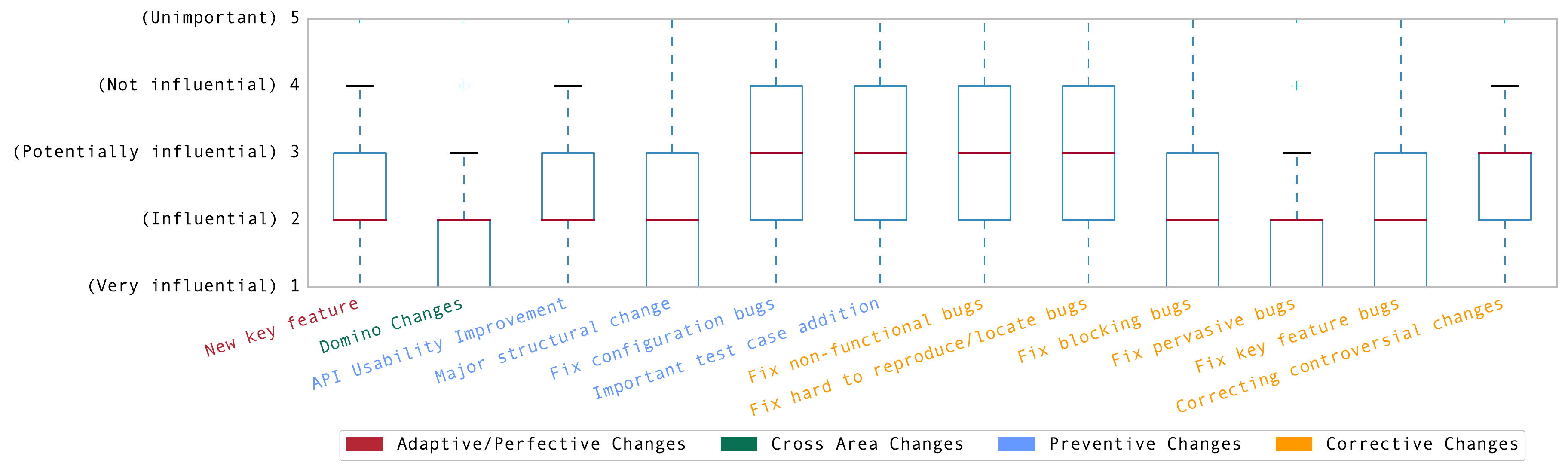}
\caption{Survey results on different categories of Influential Changes.}
\label{fig:survey}
\vspace{-0.2cm}
\end{figure*}
\subsection{Developer Validation}
\label{subsec.opencard}

To further validate the influential software changes dataset that we have
collected with our intuition-based post-mortem metrics, we perform a large-
scale developer study. Instead of asking developers to confirm each identified
commit, we must summarize the commits into categories. To that end, we
resorted to open-card sorting~\cite{Nielsen95}, a well known, reliable and
user-centered method for building a taxonomy of a system~\cite{boxesandarrows}.
Card sorting helps explore patterns on how users would expect to find content
or functionality. In our case, we use this technique to label influential
software changes within categories that are easily differentiable for
developers.

We consider open-card sorting where participants are given cards showing
description of identified influential software changes\footnote{We consider
all 177 influential software changes from the post-mortem analysis.} without
any pre-established groupings. They are then asked to sort cards into groups
that they feel are acceptable and then describe each group. We performed this
experiment in several iterations: first two authors of this paper provided
individually their group descriptions, then they met to perform another
open-card sorting with cards containing their group descriptions. Finally, a
third author, with more experience in open-card sorting, joined for a final
group open-card sorting process which yielded 12 categories of influential
software changes.

The influential software changes described in the 12 categories span over
four software maintenance categories initially defined by Lientz {\em et
al.}~\cite{Lientz:1978:CAS:359511.359522} and updated in ISO/IEC 14764. Most
influential software changes belong to the {\em corrective changes} category.
Others are either {\em preventive changes}, {\em
adaptive changes} or {\em perfective changes}. Finally, changes in one of our influential change categories
 can fall into more than one maintenance categories. We refer to
them as {\em cross area changes}.

\textbf{Developer assessment.}
We then conduct a developer survey to assess the relevance of the 12 categories of influential changes that we describe. 
The survey participants have been selected from data collected in the GHTorrent project~\cite{Gousi13} which contains history
archives on user activities and repository changes in GitHub. We consider active developers (i.e., those who have contributed in the latest changes recorded in GHTorrent) and focus on those who have submitted comments on other's commit. We consider this to be an indication of experience with code review. The study\footnote{Survey form at \url{https://goo.gl/V2g8OE}} was sent to over 1952 developer email addresses. After one week waiting period, only 800 email owners opened the mail and 144 of them visited the survey link. Finally 89 developers volunteered to participate in the survey. 66 (i.e., 74\%) of these developers hold a position in a software company or work in freelance. Nine respondents (10\%) are undergraduate students and eight (9\%) are researchers. The remaining six developers did not indicate their current situation. In total, 78\% of the participants confirmed having been involved in code review activities. 26 (29\%) developers have between one and five years experience in software development. 29 (33\%) developers have between five and ten years of experience. The remaining 34 (38\%) have over ten years of experience. 

In the survey questionnaire, developers were provided with the name of a category of influential software changes, its description and an illustrative example from our dataset. The participant was then requested to assess the relevance of this category of changes as influential software using a Likert scale between {\tt 1:very influential} and {\tt 5:unimportant}. Figure~\ref{fig:survey} summarizes the survey results. For more detailed description of the categories, we refer the reader to the project web site (see Section ``Availability'').

The survey results suggest that:
\begin{itemize}
	\item According to software developers with code review experience, all 12 categories are about important changes: 7 categories have an average agreement of 2 (i.e., Influential), the remaining 5 categories have an average of 3 (i.e., potentially influential). Some (e.g.,  ``domino changes'' and ``changes fixing pervasive bugs'') are clearly found as more influential than others (e.g., ``important test case addition'').
	\item Some changes, such as ``fixes for hard to reproduce or locate bugs'', are not as influential as one might think.
	\item Developers also suggested two other categories of influential changes: {\em Documentation changes} and {\em Design-phase changes}. The latter however are challenging to capture in source code repository artefacts, while the former are not relevant to our study which focuses on source code changes.
\end{itemize}

With this study we can increase our confidence in the dataset of influential
software changes that we have collected. We thus consider leveraging on the
code characteristics of these identified samples to identify more influential
changes.

\section{Learning to Predict IC{\small \bf s}}
\label{sec:model}

Beyond the observational study reported in Section~\ref{sec:preliminary}, we
propose an approach to identify influential changes on-the-fly. The objective
is to predict, when a change is being submitted, whether it should be
considered with care as it could be influential. To that end, the approach
leverages Machine Learning (ML) techniques. In our study, the learning process
is performed based on the dataset yielded by the systematic post-mortem
analysis and whose labels were manually confirmed. In this section we describe
the features that we use to build classifiers as well as the quantitative
assessment that was performed.

\subsection{Machine Learning Features for ICs}

A change submitted to a project repository contains information on what the
change does (in commit messages), files touched by the change, quantity of
edits performed by the change and so on. Based on the experience gathered
during manual analysis of influential changes in
Section~\ref{sec:preliminary}, we extract a number of features to feed the ML
algorithms.

\subsubsection{Structural features}
First we consider common metrics that provide hints on structural
characteristics of a change.
These metrics include (1) the number of files simultaneously changed in a single
commit, (2) the number of lines added to the program repository by the
commit and (3) the number of lines removed from the code base.

\subsubsection{Natural language terms in commit messages} During the
observational study, we noted that commit messages already contain good hints
on the importance of the change that they propose. We use the {\em Bag-of-
words}~\cite{lewis:ecml:1998} model to compute the frequency of occurrence of
words and use it as a feature. In addition, we noted that developers may be
emotional in the description of the change that they propose. Thus, we also
compute the subjectivity and polarility of commit messages based on {\em
sentiment analysis}~\cite{hu_opinion_2006,ohana_opinion_2009,
liu_sentiment_2010,thelwall_sentiment_2010} techniques.

\begin{table*}[!t]
    \caption{Performance comparison using Na\"{i}ve Bayes and Random Forest classifiers.}
    \label{tab:f-measure}
    \centering
\resizebox{\linewidth}{!}{
\begin{tabular}{c | c | c c c c c c c c | c}
& {\bf Algorithm} & {\bf Commons-codec} & {\bf Commons-collections} & {\bf Commons-compress} & {\bf Commons-csv} & {\bf Commons-io} & {\bf Commons-lang} & {\bf Commons-math} & {\bf Storm}  & {\bf Average}\\

\toprule

{\bf F-Measure}
& NB & 95.1 & 92.9 & 91.5 & 84.2 & 98.5 & 89.2 & 94.3 & 86.1 & 91.5 \\
{\bf (Influential Class)}
& RF & 97.4 & 96.4 & 98.2 & 77.8 & 97.0 & 95.0 & 99.1 & 97.8 & 94.8 \\
\midrule
{\bf F-Measure}
& NB & 93.5 & 87.5 & 83.9 & 92.7 & 98.1 & 79.5 & 92.6 & 86.5 & 89.3 \\
{\bf (Non Influential Class)}
& RF & 97.0 & 93.9 & 97.1 & 90.5 & 96.3 & 92.9 & 98.9 & 97.5 & 95.5 \\

\bottomrule

\end{tabular}
}

    \vspace{-0.3cm}
\end{table*}

\subsubsection{Co-change impact}
Finally, we consider that the frequency to which a pair of files are changed
together can be an indication of whether a given change commit affecting both
files (or not) is influential.
In our experiments, for each commit, we build a co-change graph of the entire
project taking into account the history of changes until the time of that
commit. Then, considering files that are actually touched by the
change commit, we extract common network metrics.


\textbf{PageRank}~\cite{Brin98} is a link analysis algorithm for ``measuring'' the importance of
an element, namely a page, in a hyperlinked set of documents such as the World Wide Web.
Considering a co-change graph as a linked set, we extract PageRank values for all files. When a commit change is
applied, the co-change graph is modified and PageRank values are changed. We build a feature vector taking into
account these changes on the minimum and maximum PageRank values.

\textbf{Centrality} metrics are commonly used in social network analysis to determine influential people,
or in Internet networks to identify key nodes. In our experiments, we focus on computing
{\em betweeness centrality}~\cite{freeman07}
and {\em closeness centrality}~\cite{Sabidussi66} metrics for all files associated to a commit change.
We build features by computing the deltas in the sum of centrality metrics between the
metrics computed for files involved in previous commits and for
files involved in current commit.

\subsection{Influential Change Classification}
In this section we present the parameters of our Machine Learning classification
experiments for predicting influential changes. In these experiments, we assess
the quality of our features for accurately classifying influential
changes. We perform tests with two
popular classifiers, the Na\"{i}ve Bayes~\cite{lewis:ecml:1998,mlbook}
and Random Forest~\cite{breiman_random_2001}.



In the process of our validation tests, we are interested in assessing: 1)
Whether connectivity on co-change graphs correlates with a probability
for a relevant change to be an IC; 2) If natural language information in
commit messages are indicative of ICs; 3) If structural information of changes
are indicative of ICs; 4) Whether combinations of features is best for
predicting ICs; 5) If our approach can discover ICs beyond the types of
changes discovered with post-mortem analysis.



\subsubsection{Experiment Setup}\label{sec:experiment}
To compute the feature vectors for training the classifiers, we used
a high-performance computing system~\cite{VBCG_HPCS14} to run parallel tasks for
building co-change graphs for the various project subjects. After extracting the
feature metrics, we preprocess the data and ran ten-fold cross validation
tests to measure the performance of the classification.

\textbf{Preprocessing.}
Influential software changes likely constitute a small subset of all changes
committed in the project repositories. Our manual analysis yielded very few
influential changes leading to a problem of imbalanced datasets in the training
data.
Since we try to identify influential changes, which constitute the minority
classes and learning algorithms are not adapted to imbalanced datasets, we use oversampling
techniques to adjust the class distribution. In our experiments, we leverage
the Synthetic Minority Over-sampling Techniques (SMOTE)~\cite{al.2002}.

\textbf{Evaluation Measures.}
To quantitatively evaluate the performance of our approach for predicting influential changes,
we used standard metrics in ML, namely Precision, Recall and
F-measure~\cite{mlbook,kim_classifying_2008,stat2}. {\bf Precision} quantifies the effectiveness of our machine learning-based approach to point to changes that are actually influential. {\bf Recall} on the other hand explores the capability of our approach to identify most of the influential changes in the commits set. Finally, we compute the {\bf F-measure}, the harmonic mean between Recall and Precision. We consider that both Precision and Recall are equally important and thus, they are equally weighted in the computation of F-measure.

\subsubsection{Assessment Results}
In the following paragraphs, we detail the prediction results for
influential changes using ten-fold cross validation on labelled
data. In addition, this section describes the result of influential
change prediction in the wild.

Cross validation is a common model validation in statistics to
assess how the results of a statistical analysis will generalize to an
independent data set. In machine learning experiments, it is common practice to
rely on {\em k-fold} cross validation where the test is performed {\em k} times,
each time testing on a $k^{th}$ portion of the data. We perform ten-fold cross
validation on the labelled dataset built in Section~\ref{sec:preliminary}.

In the first round of experiments, we built feature vectors with all
features considered in our study. We then built classifiers using Na\"{i}ve
Bayes and Random Forest. Table~\ref{tab:f-measure} depicts the F-measure
performance in ten-fold cross validation for the two algorithms. Although
Random Forest performs on average better than Na\"{i}ve Bayes, this difference
is relatively small.




Table~\ref{tab:ten-fold-rf} details the validation results with Random Forest for different combinations
of feature groups for the experiments.
We considered separately features relevant to co-change metrics, the natural
language commit message, and the structural information of changes. We
also combined those type of features to assess the potential performance
improvement or deterioration.

\begin{table}[h!]
    \caption{Ten fold cross validation on influential changes using
    Random Forest with different metrics combinations.
    CC: co-change features. NL: natural language terms on commit messages.
    SI: structural features.}
    \label{tab:ten-fold-rf}
    \centering
\resizebox{\linewidth}{!}{
\begin{tabular}{l l | c c c | c c c}
  \multirow{2}{*}{\bf Project Name} & 
  \multirow{2}{*}{\bf Metrics} &
  \multicolumn{3}{c|}{\bf Influential Class} & 
  \multicolumn{3}{c}{\bf Non-Influential Class} \\
  & & Precision & Recall & F-Measure & Precision & Recall & F-Measure \\

\toprule

{\tt Commons-codec}
& \highlight CC & 97.5 & 97.5 & \highlight 97.5 & 96.9 & 96.9 & 96.9 \\
& NL & 100.0 & 92.5 & 96.1 & 91.4 & 100.0 & 95.5 \\
& SI & 81.0 & 85.0 & 82.9 & 80.0 & 75.0 & 77.4 \\
& CC NL & 100.0 & 95.0 & 97.4 & 94.1 & 100.0 & 97.0 \\
& CC SI & 95.0 & 95.0 & 95.0 & 93.8 & 93.8 & 93.8 \\
& NL SI & 100.0 & 95.0 & 97.4 & 94.1 & 100.0 & 97.0 \\
& ALL & 100.0 & 95.0 & 97.4 & 94.1 & 100.0 & 97.0 \\
\midrule
{\tt Commons-}
& CC & 90.5 & 92.7 & 91.6 & 87.5 & 84.0 & 85.7 \\
{\tt collections}
& NL & 94.9 & 90.2 & 92.5 & 85.2 & 92.0 & 88.5 \\
& SI & 80.4 & 90.2 & 85.1 & 80.0 & 64.0 & 71.1 \\
& CC NL & 97.3 & 87.8 & 92.3 & 82.8 & 96.0 & 88.9 \\
& CC SI & 86.7 & 95.1 & 90.7 & 90.5 & 76.0 & 82.6 \\
& NL SI & 95.1 & 95.1 & 95.1 & 92.0 & 92.0 & 92.0 \\
& \highlight ALL & 95.2 & 97.6 & \highlight 96.4 & 95.8 & 92.0 & 93.9 \\
\midrule
{\tt Commons-}
& CC & 92.9 & 96.3 & 94.5 & 94.1 & 88.9 & 91.4 \\
{\tt compress}
& NL & 100.0 & 96.3 & 98.1 & 94.7 & 100.0 & 97.3 \\
& SI & 89.7 & 96.3 & 92.9 & 93.8 & 83.3 & 88.2 \\
& CC NL & 100.0 & 96.3 & 98.1 & 94.7 & 100.0 & 97.3 \\
& CC SI & 87.1 & 100.0 & 93.1 & 100.0 & 77.8 & 87.5 \\
& \highlight NL SI & 100.0 & 100.0 & \highlight 100.0 & 100.0 & 100.0 & 100.0 \\
& ALL & 96.4 & 100.0 & 98.2 & 100.0 & 94.4 & 97.1 \\
\midrule
{\tt Commons-csv}
& CC & 40.0 & 36.4 & 38.1 & 65.0 & 68.4 & 66.7 \\
& NL & 100.0 & 63.6 & 77.8 & 82.6 & 100.0 & 90.5 \\
& \highlight SI & 100.0 & 81.8 & \highlight 90.0 & 90.5 & 100.0 & 95.0 \\
& CC NL & 100.0 & 54.5 & 70.6 & 79.2 & 100.0 & 88.4 \\
& CC SI & 66.7 & 54.5 & 60.0 & 76.2 & 84.2 & 80.0 \\
& NL SI & 100.0 & 72.7 & 84.2 & 86.4 & 100.0 & 92.7 \\
& ALL & 100.0 & 63.6 & 77.8 & 82.6 & 100.0 & 90.5 \\
\midrule
{\tt Commons-io}
& CC & 93.9 & 93.9 & 93.9 & 92.6 & 92.6 & 92.6 \\
& \highlight NL & 100.0 & 97.0 & \highlight 98.5 & 96.4 & 100.0 & 98.2 \\
& SI & 82.5 & 100.0 & 90.4 & 100.0 & 74.1 & 85.1 \\
& \highlight CC NL & 100.0 & 97.0 & \highlight 98.5 & 96.4 & 100.0 & 98.2 \\
& CC SI & 94.1 & 97.0 & 95.5 & 96.2 & 92.6 & 94.3 \\
& \highlight NL SI & 100.0 & 97.0 & \highlight 98.5 & 96.4 & 100.0 & 98.2 \\
& ALL & 97.0 & 97.0 & 97.0 & 96.3 & 96.3 & 96.3 \\
\midrule
{\tt Commons-lang}
& CC & 86.5 & 88.9 & 87.7 & 82.6 & 79.2 & 80.9 \\
& NL & 94.4 & 93.1 & 93.7 & 89.8 & 91.7 & 90.7 \\
& SI & 72.2 & 79.2 & 75.5 & 63.4 & 54.2 & 58.4 \\
& \highlight CC NL & 95.8 & 95.8 & \highlight 95.8 & 93.8 & 93.8 & 93.8 \\
& CC SI & 91.9 & 94.4 & 93.2 & 91.3 & 87.5 & 89.4 \\
& NL SI & 98.5 & 93.1 & 95.7 & 90.4 & 97.9 & 94.0 \\
& ALL & 97.1 & 93.1 & 95.0 & 90.2 & 95.8 & 92.9 \\
\midrule
{\tt Commons-math}
& CC & 95.4 & 96.3 & 95.9 & 95.7 & 94.7 & 95.2 \\
& \highlight NL & 100.0 & 100.0 & \highlight 100.0 & 100.0 & 100.0 & 100.0 \\
& SI & 76.3 & 80.6 & 78.4 & 76.1 & 71.3 & 73.6 \\
& \highlight CC NL & 100.0 & 100.0 & \highlight 100.0 & 100.0 & 100.0 & 100.0 \\
& CC SI & 96.4 & 98.1 & 97.2 & 97.8 & 95.7 & 96.8 \\
& NL SI & 100.0 & 98.1 & 99.1 & 97.9 & 100.0 & 98.9 \\
& ALL & 100.0 & 98.1 & 99.1 & 97.9 & 100.0 & 98.9 \\
\midrule
\midrule
{\tt Spring-}
& \highlight NL & 96.2 & 90.9 & \highlight 93.5 & 84.6 & 93.2 & 88.7 \\
{\tt framework}
& SI & 75.8 & 88.2 & 81.5 & 68.3 & 47.5 & 56.0 \\
& NL SI & 96.0 & 86.4 & 90.9 & 78.6 & 93.2 & 85.3 \\
\midrule
{\tt Storm}
& CC & 97.7 & 95.5 & 96.6 & 95.1 & 97.5 & 96.2 \\
& NL & 97.8 & 98.9 & 98.3 & 98.7 & 97.5 & 98.1 \\
& SI & 90.0 & 80.9 & 85.2 & 80.7 & 89.9 & 85.0 \\
& CC NL & 97.8 & 97.8 & 97.8 & 97.5 & 97.5 & 97.5 \\
& CC SI & 97.7 & 95.5 & 96.6 & 95.1 & 97.5 & 96.2 \\
& \highlight NL SI & 98.9 & 98.9 & \highlight 98.9 & 98.7 & 98.7 & 98.7 \\
& ALL & 97.8 & 97.8 & 97.8 & 97.5 & 97.5 & 97.5 \\
\midrule
\midrule
{\tt Wildfly}
& NL & 93.7 & 98.4 & 96.0 & 98.0 & 92.6 & 95.2 \\
& SI & 78.7 & 82.3 & 80.5 & 79.0 & 75.0 & 77.0 \\
& \highlight NL SI & 96.0 & 99.2 & \highlight 97.6 & 99.0 & 95.4 & 97.2 \\

\bottomrule
\end{tabular}
}

    \vspace{-0.4cm}
\end{table}

Co-change metrics, which are the most tedious to extract (hence missing from two projects in Table~\ref{tab:ten-fold-rf} due to too large graphs)
histories, allow to yield an average performance of 87.7\% precision,
87.5\% recall, and 87.6\% F-measure.


Natural language terms in commit messages also allow to yield an average
performance of 94.9\% precision, 94.4\% recall, and 94.4\% F-measure
for the influential change class on average.


Our experiments also revealed that structural features of changes yield the
worst performance rates, although those performances reached 80.5\%
F-measure on average. For some projects, however, these metrics lead to a
performance slightly above 50\% (random baseline performance).


The performance results shown in Table~\ref{tab:ten-fold-rf} also highlight
the fact that, on average, combining different features contributes to improve
the performance of influential change prediction. Combining co-change and
natural language terms in commit messages achieves on average a precision, recall and
F-measure performance of 95.6\%, 94.5\% and 94.5\% respectively.
Similarly, combining co-change and structural features shows the F-measures at
90.1\% on average. Combinations of natural language and structural information
show 95.6\% F-measure. Finally, combining all features leads to an average
performance of 96.1\% precision, 94.9\% recall, and 95.2\% F-measure. However,
no feature combination achieves the best performance in every project, possibly
suggesting these features are specific to projects.


\subsubsection{Generalization of Influential Change Features}
In previous experiments, we have tested the machine learning classifier with
influential change data labelled based on three specific criteria (changes that
fix controversial/popular issues, isolated changes and changes referenced by other changes).
These categories are however strictly related to our initial intuitions for collecting influential
changes in a post-mortem analysis study. There are likely many influential changes that do not
fit into those categories. Our objective is thus to evaluate whether the features that we
use for classification of influential changes are still relevant in the wild.

We randomly sample a significant set of changes within our dataset of 10 projects commits.
Out of the \numChanges commits from the dataset, we randomly consider 381 commits (i.e., the
exact number provided by the Sample Size Calculator\footnote{\url{http://www.surveysystem.com/sscalc.htm}} using
95\% for the confidence level and 5 for the confidence interval).

Again we manually label the data based on the categories of influential changes approved by developers (cf. Section~\ref{subsec.opencard}).
We cross check our labels among authors and perform ten-fold cross validation using the same features presented in Section~\ref{sec:experiment} for
influential change classification. The results are presented in Table~\ref{tab:ten-fold-rf-random}.

\begin{table}[h!]
\caption{Ten-fold cross validation on randomly sampled
and then manually labelled data. We show results considering all features (NL
and SI features in the case of {\tt Spring-framework} and {\tt Wildfly}
because of missing CC features).}

\label{tab:ten-fold-rf-random}
    \centering
\resizebox{\linewidth}{!}{
\begin{tabular}{l | c c c | c c c}
  \multirow{2}{*}{\bf Project Name} & 
  \multicolumn{3}{c|}{\bf Influential Class} & 
  \multicolumn{3}{c}{\bf Non-Influential Class} \\
  & Precision & Recall & F-Measure & Precision & Recall & F-Measure \\

\toprule

{\tt Commons-codec}
& 100.0 & 88.9 & 94.1 & 87.5 & 100.0 & 93.3 \\
{\tt Commons-collections}
& 100.0 & 88.9 & 94.1 & 83.3 & 100.0 & 90.9 \\
{\tt Commons-compress}
& 0.0 & 0.0 & 0.0 & 66.7 & 100.0 & 80.0 \\
{\tt Commons-io}
& 86.7 & 86.7 & 86.7 & 75.0 & 75.0 & 75.0 \\
{\tt Commons-lang}
& 97.3 & 90.0 & 93.5 & 85.2 & 95.8 & 90.2 \\
{\tt Commons-math}
& 100.0 & 31.6 & 48.0 & 71.1 & 100.0 & 83.1 \\
{\tt Spring-framework}
& 97.5 & 96.9 & 97.2 & 91.7 & 93.2 & 92.4 \\
{\tt Storm}
& 100.0 & 88.2 & 93.8 & 88.2 & 100.0 & 93.8 \\
{\tt Wildfly}
& 100.0 & 96.4 & 98.2 & 95.8 & 100.0 & 97.8 \\

\bottomrule
\end{tabular}
}

\end{table}

The precision of ten-fold cross validation for influential changes is on average 86.8\% while the average recall
is 74\%. These results suggest that overall, the features provided in our
study are effective even in the wild. For some projects, the performance is
especially poor, mainly because 1) their training data is limited ({\tt Commons-CSV}
has only one labeled influential change, making it infeasible to even oversample, thus
no results are available in the table), 2) currently, we do not take into account some features
of influential changes related to documentation. Developers have already brought up this aspect in the survey.

\subsubsection{Evaluation Summary}



From our evaluation results we have found that: 1) Co-change metrics allow
to successfully predict influential changes with an average 87.6\% F-measure;
2) Features based on terms in commit messages can predict influential changes
with high precision (average of 94.9\%) and recall (average of 94.4\%); 3)
Structural features can be leveraged to successfully predict influential
changes with an average F-measure performance of 80.5\%; 4) Overall, combining
features often achieves a better prediction performance than individual
feature groups. For example, combining all features showed 96.1\% precision,
94.9\% recall, and 95.2\% F-measure on average; 5) With the features we
collected, our prediction approach has an acceptable performance in the wild, i.e.,
with different types of influential changes (beyond the ones we relied upon to infer
the features).

\vspace{-0.2cm}
\section{Threats to Validity}
\label{sec:discussion}
Our study raises several threats to validity. This section outlines the most salient ones.

\textbf{Internal validity.} The authors have manually labelled themselves the influential changes as it
was prohibitively costly to request labelling by a large number of developers. We have mitigated this issue
by clearly defining criteria for selecting influential changes, and by performing cross-checking.
Another threat relates to the number of developers who participated to the code developer study for
approving the categories of influential changes. We have attempted to mitigate this threat by launching
advertisement campaigns targeting thousands of developers. We have further focused on quality and representative 
developers by targeting those with some code review experience.

\textbf{External validity.} Although we considered a large dataset of commit
changes, this data may not represent the universe of real-world programs.
Indeed, the study focused on open-source software projects written in Java.
The metrics and features used for predicting influential changes in this context
may not be representative for other contexts.

\textbf{Construct validity.} Finally, we selected features based on
our intuitions on influential changes. Our study may have thus overlooked more
discriminative features. To mitigate this threat, first we have considered several
features, many of which are commonly known in the literature, second we have repeated
the experiments based on data labelled following  new category labels of influential changes
approved by developers.

\section{Related Work}
\label{sec:related}

This section discusses four groups of related work; 1) software evolution, 2)
change impact analysis, 3) defect prediction, and 4) developer expertise. These
topics address several relevant aspects of our study.

\subsection{Software Evolution}
Changing any file in a software system implies that the system evolves in a
certain direction. Many studies dealt with software evolution in different ways.
D'Ambros et al.~\cite{dambros_evolution_2006} presented \emph{the evolution
radar} that visualizes file and module-level coupling information. Although this
tool does not directly predict or analyze the change impact, it can show an
overview of coupling relationships between files and modules.
Chronos~\cite{servant_history_2012} provides a narrowed view of history
slicing for a specific file. The tool analyzes a line-level history of a file.
This reduces the time required to resolve program evolution tasks. Girba et
al.~\cite{girba_how_2005} proposed a metric called \emph{code ownership} to
illustrate how developers drive software evolution. We used the metric to
examine the influence of a change.

\subsection{Change Impact Analysis}

Many previous studies revealed a potential impact of software changes. There is a
set of techniques that use dynamic analysis to identify change impacts. Ren et
al.~\cite{ren_chianti:_2004} proposed \emph{Chianti}. This tool first runs test
cases on two subsequent program revisions (after/before a change) to figure out
atomic changes that describe behavioral differences. The authors provided a
plug-in for Eclipse, which help developers browse a change impact set of a
certain atomic change. FaultTracer~\cite{zhang_faulttracer:_2012} identifies
a change impact set by differentiating the results of test case executions on
two different revisions. This tool uses the extended call graphs to select
test cases affected by a change.

Brudaru and Zeller~\cite{brudaru_what_2008} pointed out that the long-term
impact of changes must be identified. To deal with the long-term impact, the authors proposed
a change genealogy graph, which keeps track of dependencies between
subsequent changes. Change genealogy captures addition/change/ deletion of
methods in a program. It can measure long-term impact on quality,
maintainability, and stability~\cite{herzig_capturing_2010}. 
In addition, it can reveal cause-effect chains~\cite{herzig_mining_2011} and predict
defects~\cite{herzig_predicting_2013}.

Although dynamic analysis and change genealogy can pinpoint a specific element
affected by a change in source code, its scope is limited to executed statements
by test cases. This can miss many affected elements in source code as well as
non-source code files such as build scripts and configuration settings.
Revision histories can be used for figuring out files changed frequently
together. Zimmermann et al.~\cite{zimmermann_mining_2004} first studied
co-change analysis in which the authors revealed that some files are
commonly changed together. Ying et al.~\cite{ying_predicting_2004} proposed an
approach to predicting files to change together based on revision histories.

There have been cluster-based techniques for change impact analysis. Robillard
and Dagenais~\cite{robillard_retrieving_2008} proposed an approach to building
change clusters based on revision histories. Clusters are retrieved by analyzing
program elements commonly changed together in change sets. Then, the approach
attempts to find matching clusters for a given change. The matching clusters are
regarded as the change impact of the given change. Sherriff and
Williams~\cite{sherriff_empirical_2008} presented a technique for change impact
analysis using singular value decomposition (SVD). This technique basically
figures out clusters of program elements frequently changed together. When clustering
changes, the technique performs SVD. The clusters can be used for identifying
the change impact of an incoming change.


\subsection{Defect Prediction}

Changing a program may often introduce
faults~\cite{sliwerski_hatari:_2005,kim_automatic_2006}.
Thus, fault prediction at an early stage can lead developers to achieving
a better software quality. Kim et al.~\cite{kim_predicting_2007} proposed a
cache-based model to predict whether an incoming change may introduce or not.
They used \emph{BugCache} and \emph{FixCache} that record entities and files
likely to introduce a bug and fix the bug if they are changed. The results of
their empirical study showed that the caches 46-95\% accuracy in seven open
source projects.

Machine learning classification can be used for defect prediction as well. Kim
et al.~\cite{kim_classifying_2008} presented an approach to classifying
software changes into buggy or clean ones. They used several features such as number of
lines of added/deleted code, terms in change logs, and cyclomatic complexity.
The authors conducted an empirical evaluation on 12 open source projects. The
result shows 78\% prediction accuracy on average. In addition, Shivaji et
al.~\cite{shivaji_reducing_2009} proposed a feature selection technique to
improve the prediction performance of defect prediction. Features are not
limited to metrics of source code; Jiang et al.~\cite{jiang_personalized_2013}
built a prediction model based on individual developers. Defect prediction
techniques are often faced with imbalanced datasets. Bird et
al.~\cite{bird_fair_2009} pointed out that unfair and imbalanced datasets can
lead to bias in defect prediction.

\subsection{Developer Expertise}

It is necessary to discuss developer expertise since influential changes implies
that the developer who made the changes can be influential to other developers.

As the size of open-source software projects is getting larger, developer
networks are naturally constructed and every activity in the network may affect
other developers substantially. Hong et al.~\cite{hong_understanding_2011}
reported a result of observing a developer social network. The authors
investigated Mozilla's bug tracking site to construct a developer social network
(DSN). In addition, they collected general social networks (GSNs) from ordinary
social communities such as Facebook and Amazon. This paper provides the
comparison between DSN and GSNs. Findings described in this paper include 1) DSN
does not follow power law degree distribution while GSNs do, 2) the size of
communities in DSNs is smaller than that of GSNs. This paper also reports the
result of evolution analysis on DSNs. DSNs tend to grow overtime but not much as
GSNs do.

Onoue et al.~\cite{onoue_study_2013} studied and enumerates developer activity
data in \texttt{Github.com}.
It classifies good developers, tries to understand developers, and
differentiates types of developers. However, the paper does not provide any
further implication. In addition, there is no result for role analysis and 
social structure.

Pham et al.~\cite{pham_creating_2013} reported the results of a user study which
has been conducted to reveal testing culture in OSS. The authors have
interviewed 33 developers of GitHub first and figured out the transparency of
testing behaviors. Then, an online questionnaire has been sent to 569 developers
of GitHub to find out testing strategies.

\section{Conclusion and Future Work}
\label{sec:conclusion}

In software revision histories, we can find many cases in which a few lines of
software changes can positively or negatively influence the whole project
while most changes have only a local impact. In addition, those
\emph{influential changes} can constantly affect the quality of software for a
long time. Thus, it is necessary to identify the influential changes at an
early stage to prevent project-wide quality degradation or immediately
take advantage of new software new features.

In this paper, we reported results of a post-mortem analysis on \numChanges
software changes that are systematically collected from \numSubjects open
source projects  and labelled based on key quantifiable criteria. We then
used open-card sorting to propose categories of influential changes.
After developer have validated these categories, we consider examples of
influential changes and extract features such as 
complexity and terms in change logs in order to build a prediction model.
We showed that the classification features are efficient beyond the scope
of our initial labeled data on influential changes. 
Our future work will focus on the following topics: 

\begin{itemize}
  \item Influential changes may affect the popularity of projects. We will
  investigate the correlation between influential changes and popularity metrics
  such as the number of new developers and new fork events.
  \item In our study, we used only metrics for source code. However, features of
  developers can have correlations with influential changes. We will study
  whether influential changes can make developer influential and vice versa.
  \item Once influential changes are identified, it is worth finding out who can
  benefit from the changes. Quantifying the impact of the influential changes to
  developers and users can significantly encourage further studies.
\end{itemize}

\vspace{-0.3cm}
\section*{Availability}
\makeatletter
\g@addto@macro{\UrlBreaks}{\UrlOrds}
\makeatother
We make available all our observational study results, extracted feature vectors and developer survey results in this work.
See \url{https://github.com/serval-snt-uni-lu/influential-changes}.

\balance
\vspace{-0.75\baselineskip}

\bibliographystyle{abbrv}
\bibliography{bib/ic,bib/developer}%

\balancecolumns 

\end{document}